%% file: JISA_2018.tex
\newif\ifjisa
\jisafalse

\ifjisa
	\documentclass[preprint]{elsarticle}
\else
	\documentclass{llncs}
\fi
\usepackage[utf8]{inputenc}
\usepackage{url}
\usepackage{multirow}
\usepackage{amssymb}
\usepackage{xcolor}
\usepackage{subfig}
\usepackage{enumitem}
\usepackage{rotating}

\newif\ifshowcomments
\showcommentstrue
\ifshowcomments
\newcommand{\mynote}[2]{\fbox{\bfseries\sffamily\scriptsize{#1}}
 {\small$\blacktriangleright$\textsf{\emph{#2}}$\blacktriangleleft$}}
\else
\newcommand{\mynote}[2]{}
\fi

\title{Malware triage for early identification of Advanced Persistent Threat activities}
\ifjisa
	\author[cis]{Giuseppe Laurenza\corref{cor1}}
	\ead{laurenza@diag.uniroma1.it}
	\author[cis]{Riccardo Lazzeretti}
	\ead{lazzeretti@diag.uniroma1.it}
	\author[cis]{Luca Mazzotti}
	\ead[url]{mazzotti.luca.93@gmail.com }
	
	\cortext[cor1]{Corresponding author} 
	\address[cis]{Research Center of Cyber Intelligence and Information Security (CIS), \\
		Dept. of Computer, Control, and Management Engineering Antonio Ruberti, Sapienza University of Rome, IT}
\else
	\author{Giuseppe Laurenza\inst{1}\orcidID{0000-0002-3763-4598} \and
	Riccardo Lazzeretti\inst{1}\orcidID{0000-0003-3835-9679} \and
	Luca Mazzotti\inst{1}}
\authorrunning{G. Laurenza et al.}
\institute{Research Center of Cyber Intelligence and Information Security (CIS), \\
	Dept. of Computer, Control, and Management Engineering Antonio Ruberti, Sapienza University of Rome, IT\\
	\email{\{laurenza,lazzeretti\}@diag.uniroma1.it, mazzotti.luca.93@gmail.com}
}

\fi
    
\begin{document}

\ifjisa
\else
	\maketitle
\fi
\begin{abstract}
    In the last decade, a new class of cyber-threats has emerged. 
    This new cybersecurity adversary is known with the name of ``Advanced Persistent Threat'' (APT) and is referred to different organizations that in the last years have been ``in the center of the eye'' due to multiple dangerous and effective attacks targeting financial and politic, news headlines, embassies, critical infrastructures, TV programs, etc. 
    %
	In order to early identify APT related malware, a semi-automatic approach for malware samples analysis is needed.  In our previous work we introduced a \emph{malware triage} step for a semi-automatic malware analysis architecture. This step has the duty to analyze as fast as possible new incoming samples and to immediately dispatch the ones that deserve a deeper analysis, among all the malware delivered per day in the cyber-space, the ones that really 
	worth to be further examined by analysts.
    Our paper focuses on malware developed by APTs, and we build our knowledge base, used in the triage, on known APTs obtained from publicly available reports. In order to have the triage as fast as possible, we only rely on static malware features, that can be extracted with negligible delay, and use machine learning techniques for the identification. In this work we move from multiclass classification to a group of oneclass classifier, which simplify the training and allows higher modularity.     
     The results of the proposed framework highlight high performances, reaching a precision of 100\% and an accuracy over 95\%. 
\end{abstract}

\input{tex/introduction}
\input{tex/related_works}
\input{tex/tools}
\input{tex/methodology}
\input{tex/testsandresults}

\input{tex/conclusion}
\ifjisa
\else
	\bibliographystyle{splncs04}
\fi
\bibliography{full_references}
\end{document}

%% file: tex/introduction.tex
\section{Introduction}\label{s:intro}
Since the origin of Internet, cyber-attacks have evolved in parallel with computer development, changing ways and means of execution. 
From the first viruses and worms to the modern botnet and rootkit, from the first singular and not-organized attack to the advanced and well-crafted persistent one, all these kinds of vector mutate their behavior updating their ``source'' in line with the new software and hardware technologies. During the last two decades, the number of delivered malware exponentially increased: according to a survey conducted by Panda Security \cite{panda}, just in 2015 about  230.000 malware were delivered per day with an increase of 40\% with respect to 2014, kneeling down different times the cybersecurity community. Recent analysis from McAfee~\cite{mcafee} calculated that in the last quarter of 2017, there were discovered more than 60 millions of new malware.
%
In the mid-2000s, indeed, the black hat community evolved from adolescent hackers to organized crime networks, fueling highly profitable identity theft schemes with massive loads of personal data harvested from corporate and government networks. In recent times, in fact, a new powerful and dangerous threat is on the rise, identified by the community as ``Advanced Persistent Threat'' (APT). 
%
%
%
According to NIST Glossary of Key Information Security Terms\footnote{http://nvlpubs.nist.gov/nistpubs/ir/2013/NIST.IR.7298r2.pdf}, APT is ``\emph{an adversary that possesses sophisticated levels of expertise and significant resources which allow it to create opportunities to achieve its objectives by using multiple attack vectors (e.g., cyber, physical and deception)}''. Hence the APT name identifies the main peculiarities of the threat:
\begin{description}[nosep,noitemsep]
\item[{Advanced}] Criminal minds behind attacks utilize the full spectrum of computer intrusion technologies and techniques. While individual attacker may not be classed as particularly ``advanced'' (e.g.  single stage malware component found on the black market), their operators typically access and develop more advanced tools as required. 
\item[{Persistent}] Criminal operators give priority to a specific task, rather than opportunistically seeking immediate financial gain. The attack is indeed conducted through continuous monitoring and interaction in order to achieve the defined objectives. A ``low-and-slow'' approach is usually more successful. 
\item[{Threat}] The attack has a malicious nature. Malevolent attacker have a specific objective and are skilled, motivated, organized and of course well funded. 
\end{description}
Moreover this advanced attacks are strongly targeted in order to overcome all the \emph{general} defences that one can apply.

According to the FireEye studies \cite{fire}, based on the huge amount of APT operations they analyzed
, APTs principally target big companies, critical infrastructure and institutions, in order to gain financial secrets, intellectual properties
, national secret, private personal information or damage critical infrastructure by interrupting or decreasing their functionality.
Despite different targets and origin, it has been demonstrated that APTs perform attacks in 
a standard way, that can be represented as an intrusion kill chain \cite{hutchins2011intelligence}. \textit{Reconnaissance} phase implies APT to define a clear understanding about the overall victim's infrastructure, in order to hence perform the \textit{initial exploitation} phase, where those malicious organizations actually ``enter'' inside the victim systems. Given the persistent appellation which characterizes APT, the next phase involves \textit{establish persistence} in the host which provides attackers the possibility to install backdoor or dropper for following attack stages. \textit{Lateral movement} is actually accomplished to escalate privileges and hence being able to elevate the capability in the system. Finally, in the \textit{exfiltration} phase the real aim of the attack is achieved.
It is hence important to identify any malware used by APTs in any of these phases, so that other malware used in previous steps can be find and analyzed and future activities can be synthesized.

\subsection{Contribution}
It is evident that there is the need for a prioritization mechanism in order to promptly identify samples that deserve to be further analyzed by secure analyst. To this end, in this paper we propose a malware triage stage, where samples are timely analyzed to understand as soon as possible whether they likely belong to some known APT campaign and should be dispatched, with highest priority, to human analysts for further analysis. 
We underline that the objective of this triage stage is not APT malware detection, which is instead pursued at a later stage by human analysts and specialized architecture components, rather the final goal of the triage is spotting with the highest precision samples that seem to be related to known APTs. There is hence the necessity of a triage stage that has low computational complexity to timely analyze a great number of samples, and provides at the same time high precision and accuracy. We privilege precision to accuracy, in fact precision is necessary because the malware triage should not prioritize non-APT malware (\textit{false positive}), to not overload human analysts and/or complex components with urgent but not necessary analyses. However also, high accuracy is desired, with the aim of correctly identify that samples really belonging to some known APT classes (\textit{true positive}). 

In our previous work \cite{laurenza2017malware}, we proposed an approach based on a \emph{Random Forest Classifier}, trained on a knowledge base built upon a collection of reports about APTs publicly released by cyber-security companies. The classifier, given the set of features obtained from static analysis, outputs the APT class the malware sample belongs to, or a ``non-APT label'', in the case the sample is not associated to any class. Such solution presents two problems. First of all, this so called \emph{negative class} has a cardinality of several orders of magnitude bigger than the set of APT samples and there is a large variety among its samples. We approached the problem by adding samples not related to APTs in the training phase, but the number of used samples cannot be really representative of the whole class. 
Finally, the solution works well and achieves great accuracy and precision, but has the disadvantage of requiring a long training time and needs to be retrained each time new APT samples or even a new APT class  are discovered.

To overcome such problems, in this paper we presents several key improvements w.r.t. existing literature:
\begin{itemize}[nosep,noitemsep]
	\item We propose a novel  modular and lightweight malware triage framework, based on \textit{One-class classification} which permits to train each classifiers only on samples related to the relative class. This permits to add new APT classifiers and modify a single classifier when new samples are discovered, without affecting the whole triage framework;
	\item We design the malware triage framework on the \textit{Isolation Forest} learning concept. An {Isolation Forest} is hence trained on the samples of each class;
	\item We train the classifier with more than 2000 samples belonging to 19 different APTs: \textit{Patchwork, APT29, Winnti Group, Lazarus Group, Carbanak, Volatile Cedar, NSA, Transparent Tribe, Shiqiang, Roaming Tiger, Violin Panda, Sandwork, Hurricane Panda, APT30, Mirage, Desert Falcon, Molerats, APT28} and \textit{Lotus Blossom};
	\item We introduce features dimensionality reduction through \textit{Linear Discriminant Analysis} in order to decrease the computational time needed in the training phase and increase the overall precision and accuracy for each class;
	\item We show that such approach allows fast, precise and also accurate identification of APTs malware, through an experimental evaluation performed on a dataset composed by samples obtained by public APT reports. We also test its performance on a non-APT dataset.
\end{itemize}

%

%% file: tex/related_works.tex
\section{Related Works}\label{s:related_work}
The awareness around APT is increasing in the last years, becoming an important  research topic within the cybersecurity area. Important works have been done to deploy APT detection as well as avoidance frameworks in order to identify compromised hosts.
In \cite{laurenza2016architecture}, authors propose the design of an architecture composed by various tangled phases that starts from the uninterrupted
collection of malware and reports from different sources, continues through different analysis components, that work on single element (like static analysis tools) and on group of element (like classification algorithms) and at the end stores all the information in a knowledge base that can be easily linked to other in order to share different information. Inside such architecture several malware analysis tools can be used. 
The number of works carried out to identify and classify malware samples, even if not focused on APT, is really huge, hence we here present a small subset of them, more related to our work. In Section ~\ref{sec:malware_id}, we focus on malware analysis through machine learning techniques, while in Section \ref{sec:APT_id}, we present works related to APT malware.

\subsection{Malware identification}\label{sec:malware_id}
An important malware analysis technique is the BitShred framework \cite{bitshred}, which extracts information from the sample and, using feature hashing, creates probabilistic data structure in order to large-scale correlate samples.
Another fast and precise malware analysis framework proposed in \cite{kirat2013sigmal} is SigMal, a framework improving the state-of-the-art of the previous systems based on the concept of malware similarity by leveraging signal processing techniques to extract noise-resistant signatures from the samples. 
In \cite{makandar2015malware}, malware triage has been deployed by machine learning classification using Artificial Neural Network. In particular, the dataset used in the classification involves 3131 samples spread over 24 different unique malware classes and the overall accuracy in detecting and classifying them is about 96\%. 
Each binary executable is represented as a greyscale image through GIST descriptors \cite{oliva2005gist}; then neural networks are used to derive and extract similar patterns among the various samples.
In \cite{ahmadi2016novel}, Ahmadi et al. propose a learning-based system using different malware characteristics to assign malware samples to their corresponding families. Because of accurate and fast classification, they propose to extract static features both from the binary hex view and the assembly one to exploit complementary information by these two representations, without relying on other information derivable from dynamic analysis of malware. 
Kong et al. \cite{kong2013discriminant} propose a framework for automated malware classification based on function call graph of the malware. First, they disassemble the single sample building the relative function call graph. Then, discriminant malware distance is computed on each pair of samples: they perform pairwise graph matching between the attributed function call graphs of two malware instances in order to measure their structural similarity. 
In \cite{nari2013automated}, Nari and Ghorbani introduce a framework for automated classification of malware samples based on their network behavior. IP addresses, port numbers, protocols and dependencies between network activities are abstracted for further analysis. 

\subsection{APT detection}\label{sec:APT_id}
Most works about APT detection focus on the identification of indicators of suspicious activities, like anomaly detection systems. In \cite{friedberg}, authors propose an anomaly detection system for APTs built around several security logs caught. In particular, this white-list based approach, keeps track of system events, their dependencies and occurrences, learning the normal system behavior and reporting all \emph{strange} actions. Another interesting work is the framework proposed by Marchetti et al. \cite{marchetti} that aims to detect, among all the hosts inside the company, the ones infected by APT in order to be further analyzed in the future. 
In line with the idea of the previous works, Ussath et al. \cite{ussath2015concept} develop a Security Investigation Framework (SIF)  that supports the analysis and the tracing of multi-stage APTs. In particular, it leverages different information sources, in order to give a comprehensive overview of the different stages of an attack.

Our previous work \cite{laurenza2017malware} aims to recognize malware developed by APTs, but contrary to the other described works, we focus on the sample analysis instead of using the malware activity during the infection. The work achieves an high precision and a very high accuracy in detecting the ownership of these kind of malware, but suffers the problems highlighted in Section \ref{s:intro}.

%% file: tex/tools.tex
\section{Tools}\label{sec:tools}
In this section, we introduce the main tools used in this paper. In particular, Section \ref{section2} presenta on PEFrame, a static malware analysis tool we use for feature extraction, while we outline Random Forest and Isolation Forest in Sections \ref{section4} and \ref{section5} respectively. Section \ref{section6} presents principles of feature reduction.

\subsection{PEFrame}\label{section2}
PEFrame\footnote{\url{https://github.com/guelfoweb/peframe}} is an open source tool that performs static analysis on 
Portable Executable malware and generic suspicious file. The Portable 
Executable format contains the information necessary for the Windows OS 
loader to manage the wrapped executable code. 
It consists of the MS-DOS Stub, the PE file Header, and the sections, and can provide an enormous amount of features, containing 
relevant information for a malware analyst.
In particular, in our work, we consider seven feature classes to be extracted from the PE, here presented:
%
%
%
%
%
%
\begin{description}[nosep,noitemsep]
\item[{Optional Header} (30 features)]   Every file has an optional header that provides information to the loader. This header is optional in the sense that some files (specifically, object files) do not have it. For image files, this header is required. An object file can have an optional header, but generally this header has no function in an object file except to increase its size. These features are extracted from the optional header of the PE. It contains information about the logical layout of the PE file, such as the address of the entry point, the alignment of sections and the sizes of part of the file in memory;
\item[{MS-DOS Header} (17 features)] The MS-DOS executable-file header contains four distinct parts: a collection of header information (such as the signature word, the file size, etc.), a reserved section, a pointer to a Windows header (if one exists), and a stub program. 
MS-DOS uses the stub program to display a message if Windows has not been loaded when the user attempts to run a program. In this contest, we are interested in features related to the execution of the file, including the number of bytes in the last page of the file, the number of pages or the starting address of the \textit{Relocation Table};
\item[File Header (7 features)] The Windows executable-file header contains information that the loader requires for segmented executable files. This  includes the linker version number, data specified by the linker, data specified by the resource compiler, tables of segment data and tables of resource data. Moreover, the features related to this class highlights information about timestamp and the CPU platform which the PE is intended for.
\item [{Obfuscated String Statistics} (3 features)] Malware authors encode strings in 
their programs to hide malicious capabilities and impede reverse engineering. 
Even simple encoding schemes defeat the ``Strings'' tool and complicate static 
and dynamic analysis. In addition to PEFrame, we use functionalities of the 
FireEye Labs Obfuscated String Solver (\textit{FLOSS}\footnote{\url{https://github.com/fireeye/flare-floss}}). It is an 
open source tool that automatically detects, extracts and decodes obfuscated 
strings, such as malicious domains, IP addresses, suspicious file paths, etc.from  Windows Portable Executable files availing of advanced static 
analysis techniques. We use this tool to 
compute some statistics, like how many entry-points or relocations are present 
in the file. 
\item[{Imports Features} (158 features)]  Following the Header are the actual sections of the file, each of which contains useful information. In particular, the \textit{.rdata} section typically contains import and export information. Sometimes there is the possibility for a double section existing, one relative to the import function (\textit{.idata}), and the other relative to the export one  (\textit{.edata}). Functions can be imported from other executables or from DLLs. We are interested in the import of a specific set of known DLLs and APIs, and use their occurrences as feature. We also use three counters representing the total number of imported APIs, the total number of imported DLLs, and the total number of exported functions. 
\item[Function lengths (50 features)]  FLOSS also provides measurements of function lengths. This class contains different counters to store that information. Due to the huge number of different functions, we use \textit{bucketing} to reduce the number of possible features. 
\item[{Directories} (65 features)] The data directory indicates where to find other 
important components of the executable information in the file. It is simply an 
array located at the end of the optional header providing information regarding 
addresses and sizes about the different data structure of the PE file. 
Similarly to what we do for imports, we check the occurrence in the file of 
some particular directory names, using their size as features. 
\end{description}

\subsection{Random Forest Classifier}\label{section4}
A Random Forest \cite{breiman2001random} is a classification tool that aggregates the results provided by a set of 
\textit{Decision Trees} through a \textit{Bootstrap aggregated} (a.k.a. \textit{Bagging}) technique. 

In general, a \emph{Decision Tree} \cite{grajski1986classification} is a decision support tool that uses a 
tree-like graph as 
model of decisions.
In particular, it is a map of the possible 
outcomes of a series of related choices, allowing a user to weigh possible 
actions against one another based on their costs, probabilities and benefits. 
In machine learning science, 
\textbf{Decision tree learning} concept uses decision trees as a predictive 
model to go from observations about  an object to conclusion about the 
objects' target value, represented in the leaves of the tree. 

Although single decision trees can be effective classifiers, increased accuracy 
often can be achieved by combining the results of a collection of decision 
trees. This is indeed a well-known procedure called \textbf{Bootstrap 
Aggregation} or \textbf{Bagging} that reduces variance combining the predictions from multiple 
high-variance decision trees to make 
more accurate predictions that any individual model. 

Random Forest Classifier is a supervised classification algorithm.
As highlighted by the name, it creates a forest by randomly ensembling different decision trees. The ensemble 
method for random forest is focused on \textit{feature bagging} concept, which has the advantage of 
significantly decreasing the correlation between each decision tree and thus 
increasing, on average, its predictive accuracy. Feature bagging works by 
randomly selecting a subset of the feature dimensions at each split in the 
growth of the individual decision tress. Although this might sound 
counterproductive since it is often desired to include as many features as 
possible, it has the purposes of deliberately avoiding on average very strong 
predictive features that lead to similar splits in trees, thereby increasing 
correlation. If a particular feature is strong in predicting the response 
value, then it will be selected for many trees.  

 To classify a new object from an input vector, the 
algorithm uses the 
input vector into each of the trees composing the forest. Each tree gives a 
``vote'' 
classification for that class, and the forest chooses the classification having 
the highest number of votes among all the trees in the forest. 

\subsection{Isolation Forest Classifier}\label{section5}
In classification problems, there is the possibility to rely on \textbf{one-class classification}, where samples used in training phase belong to the same class and in the classification phase we use the classifier to identify outliers, i.e. sample not belonging to the class used for training.

Isolation forest \cite{liu2008isolation} is a one-class classifier, which data structure is built on the same conceptual principle of 
Random forest. In particular, an Isolation forest is still an ensemble of 
bootstrapped decision trees, but, instead of profiling normal points, it 
outcomes all possible anomalies that are \textit{isolated} with respect to the 
models just created. 
As stated by Zhou et al. \cite{liu2008isolation}, anomalies are data 
patterns that have different data characteristics from normal instances. Many 
existing model-based approaches to anomaly detection construct a profile of 
normal instances, and hence identify instances not conform to the normal 
profile as isolated. 

Technically speaking, isolation forest algorithm isolates observation by 
randomly selecting a feature and then randomly selecting a split value between 
the maximum and minimum values of the selected features. Hence, the algorithm 
first constructs the separation by creating isolation trees or random decision 
trees; then, when they collectively produce shorter path lengths for some 
particular points, they are highly likely to be anomalies. 
Based on the average path length derived in testing phase, it calculates the 
\textit{anomaly score} for each sample. In the various online tool libraries, 
it is possible to set a \textit{tolerance} threshold that discriminates the 
given sample as isolated or not according to the model: the higher it is, more 
tolerant the isolation forest is with anomalies.

\subsection{Principles of features reduction}\label{section6}
When dealing with high-dimensional features  space, 
it is actually infeasible to think that each feature has the same 
``importance''. 
In particular, both \textit{redundant} and \textit{irrelevant} 
features exist. 
%
Feature 
reduction techniques can be hence used to reduce the dataset dimensionality in order to facilitate training process, decrease training and classification computational time, reduce the variance among features, and avoid  \textit{curse of dimensionality}, i.e. with the increasing of 
the dimensionality, space volume increases so fast that training data become 
sparse.

In this paper, for features reduction 
purpose, we rely on the \textbf{Linear Discriminant Analysis} (LDA) technique, because of its simplicity. LDA is a generalization of Fisher's linear discriminant \cite{welling2005fisher}, a method used in statistics, pattern recognition and machine learning to project the feature space in a smaller feature space through a linear combination of features. 
In practice, given the original dimension $M$ of the features, it is possible to reduce the dimension to a chosen $L$ by 
projecting into the linear subspace $H_{L}$ maximizing inter-class 
variance after projection.

%% file: tex/methodology.tex
\section{Methodology}\label{s:methodology}
In this section, we define the methodologies that lead us to define the proposed framework with high performances. In particular, in Section \ref{extra}, we  explain the feature extraction process, in Sections \ref{sec:multi_id} and \ref{sec:one}, we respectively discuss Multi-Class methodologies proposed in \cite{laurenza2017malware} by Baldoni et al. w.r.t. the One-Class classification approach through which we implement the malware triage. 

\subsection{Features Extraction}\label{extra}
We have built our knowledge base  by
crawling external publicly-available sources to collect reports related to malicious campaigns, activities and software associated to APTs. These reports are produced by security firms and contain different \textit{Indicator of Compromises} (IoCs) related to specific APTs, including domains, IPs and MD5 of malware. 
APTnotes\footnote{\url{https://github.com/aptnotes}}
provides a Coma-Separated Values (\textit{.csv}) containing more than 400 APT reports, with \textit{Title, Source, Link, Date} and \textit{SHA1} for each file. Reports contain a rich description relative to known APT campaigns, showing details of the attacks such as the vulnerabilities exploited, infecting vectors, C\&C domains and IPs and, in most cases, at least one MD5 of the executable involved in the attack. Since the supervised nature of the inner framework classifier, we have to discriminate in each report a relative APT class, in order to correctly label each of the available source. Hence, with IoC Parser tool\footnote{\url{https://github.com/armbues/ioc_parser}}, we extrapolate hashes in the file in order to crawl for the correlative binary. As shown in Table \ref{tab:aptclass}, we come up with more than 2,000 samples related to nineteen distinct classes. 
\begin{table}[!t]
	\caption{APTs elements per classes}	\label{tab:aptclass}
	\begin{minipage}{.5\linewidth}
		\centering
		\begin{tabular}{|l|r|}
			\hline
			\textbf{Class}    & \multicolumn{1}{l|}{\textbf{Count}} \\ \hline
			Patchwork         & 559                                 \\ \hline
			APT29             & 205                                 \\ \hline
			Winnti Group      & 176                                 \\ \hline
			Carbanak          & 105                                 \\ \hline
			Volatile Cedar    & 35                                  \\ \hline
			NSA               & 13                                  \\ \hline
			Transparent Tribe & 267                                 \\ \hline
			Shiqiang          & 31                                  \\ \hline
			Roaming Tiger     & 14                                  \\ \hline
			Mirage            & 54                                  \\ \hline
		\end{tabular}
	\end{minipage}%
	\begin{minipage}{.5\linewidth}
		\centering
		\begin{tabular}{|l|r|}
			\hline
			\textbf{Class}  & \multicolumn{1}{l|}{\textbf{Count}} \\ \hline
			Lazarus Group   & 58                                  \\ \hline
			Sandwork        & 44                                  \\ \hline
			Hurricane Panda & 315                                 \\ \hline
			APT30           & 101                                 \\ \hline
			Violin Panda    & 23                                  \\ \hline
			Desert Falcon   & 45                                  \\ \hline
			Molerats        & 25                                  \\ \hline
			APT28           & 68                                  \\ \hline
			Lotus Blossom   & 48                                  \\ \hline
		\end{tabular}
	\end{minipage}
\end{table}
Retrieved the malware samples, we process them with the PEFrame tool to extract features from each binary. As highlighted in Section \ref{section2}, for each analyzed executable, the extraction tool provides us seven distinct features classes with more than 300 characteristics. 

\subsection{Multi-class classification}\label{sec:multi_id}
The framework proposed by Baldoni et al. \cite{laurenza2017malware} creates multi-class model in order to deal with APT classification. In particular, they rely on Random-Forest classifier, using samples related to known APTs for training. They also consider for training phase an additional class to represent all the samples that have not been created by any known APT. However, the elements constituent this latter class are not homogeneous among them: if it is possible to define a classification model for any APT class since the similarity among elements produced by the same organization, the same concept is not applicable for malware not related to APT. Moreover, the inequality amount of samples belonging to this class with the respect to other classes leads to an excessive imbalance in the training set. For this reason, in training phase, authors only consider classes of known APT. In order to discriminate a sample as not related to any APT, they produce a threshold for each class in the Knowledge Base (KB): if a given sample analyzed is under each APT threshold, it is considered to represent a non-APT malware. 

Baldoni et al. tested the algorithm with $10$-fold cross validation, a common value in literature. For each execution, they generate the model with $k-1$ folds and test it with both the remaining fold and all the collected malware not developed by APTs. 


\subsection{One-class classification}\label{sec:one}
In the multi-class classification, each single APT sample that is going to enforce the Knowledge Base needs a total re-processing phase for Random Forest in order to reproduce each APT threshold, wasting time and resources. 
%
Instead of relying on multi-class classification, indeed, we are going to generate, for each single APT class, an Isolation Forest, that creates a model that identifies all those elements belonging to it while discarding all anomalies. If a given sample passes ``unnoticed'' through each APT classifier, it is considered not related to any known APT. When a single known APT aimed to enforce our knowledge base is identified, we simply re-process training phase for the related APT classifier. 

We implemented our classifiers on the open-source Scikit library\footnote{\url{http://scikit-learn.org/}}, that provides different machine learning tools in Python programming language. 
Linear discriminant analysis has been used to reduce the feature dimensionality from 300 to 18 features\footnote{We decided to relate the number of projection to the number of APTs considered, however other values should be investigated.}. After having split the dataset in training set and validation set through 10-folder cross-validation, we have trained one isolation forest for each class.
Isolation Forest implementation requires two main arguments to tune the fitting phase: \textit{contamination} expresses how much the classifier has to be tolerant in creating the model for detecting outliers; \textit{number of estimators} is an attribute totally related to the tree-nature of the Isolation Forest and it defines how many trees it has to use in the ensemble methods. Despite the training set of each class is not contaminated by outliers, malware samples have high feature variability, hence we decided to provide a non-zero value to contamination. This means that some of the training samples can be wrongly considered as outliers, but increases precision in the classification phase. In order to tune the $\langle$\textit{contamination}, \textit{number of estimators}$\rangle$ tuple for achieving the best result for each Isolation Forest classifier, we perform cross-validation over the APT training set to generate confusion matrix for each APT class and the relative performances of the classifiers. 
%

In the choice of the correct parameters, we mainly rely on i) \textit{Precision}, since the malware triage nature of the proposed framework. Facing with an high value of false positives is deleterious due to the fact that human analysts are limited resources and the triage approach must reduce as much as possible their workload by striving to deliver them only samples that are highly confidently related to known APTs; ii) \textit{Accuracy}, expressing how much the APT classifier is accurate in determining the right APT class of an APT sample.
The choice of the contamination-number of estimators parameters for the Isolation Forest is hence totally based on precision-accuracy trade-off. Since the triage nature of the framework, we have been more oriented to tune this diatribe toward precision measure, due to stating a given sample belongs or not to a given APT is more effective, in a triage,  than declaring that an APT sample belongs to a class. 

%% file: tex/testsandresults.tex
\section{Analysis}\label{s:experiment}
To show that also this new approach can achieve good results we evaluate it with in different ways. In Section \ref{sec:test}, we consider only the output of the algorithm as \emph{belonging} or \emph{not belonging} to any APTs, in order to test and measure the performances of the triage work that discriminates 
APT samples from the non-APT ones. In Section \ref{sec:identification}, we identify, among the samples previously labeled to belong to some APT, the relative APT class, 
showing the achieved results for each classifier. Finally, in Section \ref{sec:comparison}, we compare our work with some of the aforementioned frameworks presented in Section \ref{s:related_work}.

A draft of the code with the used dataset can be found on GitHub\footnote{\url{https://github.com/GiuseppeLaurenza/I_F_Identifier}}

\subsection{APT-Triage}\label{sec:test} 

In order to generate a clear and uniformed test, we  rely on \textit{Stratified 10-Fold Validation}, available on Scikit. 
In particular, it splits data in train/test sets, preserving the percentage of samples for each class. 
For each APT class in the dataset, we train an Isolation Forest with the samples related to that class, thus creating 19 Isolation Forests. A sample is recognized to belong to some APT if at least one of the Isolation Forest outputs it belongs to its class. Contrarily, a malware is classified as \textit{non-APT} if no one of the classifiers recognizes it.   
 
With the priori knowledge of which element really belongs to APT or non-APT class, it has been possible to plot a global confusion matrix that shows how many samples have correctly been labeled.

\begin{table}[b]
	\centering
\caption{APT/non-APT Confusion Matrix.}
\label{tableCM}
	\subfloat[][Confusion matrix]{%
\begin{tabular}{ll|r|r|}
\cline{3-4}
                                                     &                                    & \multicolumn{2}{c|}{\textit{Predicted}}                                     \\ \cline{3-4} 
                                                     &                                    & \multicolumn{1}{l|}{\textbf{non-APTs}} & \multicolumn{1}{l|}{\textbf{APTs}} \\ \hline
\multicolumn{1}{|c|}{
	\multirow{2}{*}{\textit{True}}} & \textbf{non-APTs}                  & 8620                                   & 0                                  \\ \cline{2-4} 
\multicolumn{1}{|c|}{}                               & \multicolumn{1}{c|}{\textbf{APTs}} & 495                                    & 1711                               \\ \hline
\end{tabular}
}
\hspace{0.5cm}
\subfloat[][Performance measures]{%
\begin{tabular}{ll@{\quad}ll}%
	\textbf{Accuracy:} & \textit{95\%}\\%
	\textbf{Precision:} &\textit{100\%}\\%
	\textbf{Recall:} &\textit{77\%}\\%
	\textbf{F1-Score:} &\textit{87\%}%
\end{tabular}%
}%
\end{table}

Table \ref{tableCM} shows the results obtained in term of quality measures: we achieve our goal of maximize the precision, in fact we obtain a 100\% precision, meaning that non-APT samples have never been identified as APT, and an accuracy of 95\%. Our system do not achieve a very high recall, but this was an expected result and it is not a real problem, due to the fact that the goal of this preliminary phase in a complex architecture is the detection of malware that belongs to APT with high probability. However we remind that  APT samples not recognized by any Isolation Forest are not excluded by further analysis of the security expert: simply they are not prioritized to a prompt and deep analysis.

\subsection{APT identification}\label{sec:identification} 
We here evaluate the ability of the trained isolation forests to exactly identify the APT the sample belongs to.  Hence, we have analyzed the performances of each classifier in the labeling process of its relative samples, over all samples that in the triage have been labeled as APT members. 
We provide global confusion matrix and metrics in Table \ref{tab:globa}, calculated as the sum of each APT confusion matrixes. In the table related APT identify the APT the isolation forest is trained to while other APT identifies the set of samples belonging to other APTS. non-APT samples are not involved in this analysis.

The overall results, presented in Table \ref{tab:globa}, points out an accuracy up to 98\% and a precision of 86\%. 
The proposed framework shows high performances in almost each APT classifier, in particular for accuracy measure: hence, if a previous classifier recognizes a determined sample to belong to its model, we are confident this identification is truthful.  
When dealing with Advanced Persistent Threat knowledge base construction, however, it can exist an inner imprecision in the discrimination between distinct APTs, derived by the possible in-common implementation or use of the same tools (e.g. \textit{Poison Ivy} \cite{bennett2013poison}, a known toolset used by different APTs). Thus, reports describing operations of a particular threat might contain MD5 also related to another APT, impacting on the classifier prediction and producing possible true-negative.

\begin{table}[t]
\centering
\caption{Global APT Results}%
\label{tab:globa}
\subfloat[][Confusion matrix]{%
\begin{tabular}{ll|r|r|}
\cline{3-4}
                                                     &                                    & \multicolumn{2}{c|}{\textit{Predicted}}                                     \\ \cline{3-4} 
                                                     &                                    & \multicolumn{1}{l|}{\textbf{Other APTs}} & \multicolumn{1}{l|}{\textbf{Related APT}} \\ \hline
\multicolumn{1}{|c|}{\multirow{2}{*}{\textit{True}}} & \textbf{Other APTs}                  & 41475                              & 249                                  \\ \cline{2-4} 
\multicolumn{1}{|c|}{}                               & \multicolumn{1}{c|}{\textbf{Related APT}} & 533                                    & 1663                               \\ \hline
\end{tabular}
}
\hspace{0.5cm}
\subfloat[][Performance measures]{%
\begin{tabular}{ll@{\quad}ll}%
\textbf{Accuracy:} & \textit{98\%}\\%
\textbf{Precision:} &\textit{86\%}\\%
\textbf{Recall:} &\textit{75\%}\\%
\textbf{F1-Score:} &\textit{80\%}%
\end{tabular}%
}%
\end{table}

\subsection{Comparison with the State of the Art}\label{sec:comparison}
Both the multi-class classification approach proposed in \cite{laurenza2017malware} and our one-class technique reach encouraging results. In particular, the framework proposed by Baldoni et al. shows a precision of 100\% and an accuracy of 96\% in the detection of APT samples, while, in our work, we still have a precision of 100\%, and an accuracy up to 95\%. However, tests in \cite{laurenza2017malware} have been performed on a smaller number of APT sets, presenting low variability in their samples. Moreover, as already discussed in Section \ref{sec:one}, the structure of our framework leads to a performant modularity, allowing to quickly train each single classifier without wasting time and resources in the calculation of the threshold for each single class.

In \cite{nari2013automated}, the automated framework proposed by Nari and Ghorbani aims to classify malware samples with analysis of network behavior. They use more than 2907 malware samples distinguished among 13 malware families, reaching an accuracy of 94\%. We outperform these results reaching an accuracy up to 98\% and a precision of 100\% on the 19 APT class considered. 

In \cite{friedberg}, an anomaly detection system for APTs is built around several security logs caught. The objective of this work is related to alert the security analysts in case an activity deviates from the normal behavior defined by ADS. However, the powerful, effectiveness and system-adaptive campaigns acted by APTs lead these approaches to not be able to dynamically react to these threats.  

Rieck et al. \cite{rieck2008learning} exploit SVM classifiers for constructing models able to classify 14 of the most common Windows malware families. In particular, they rely on the analysis of dynamic features extrapolated by supervising malware behavior in a safe environment. In addition to the slowness in creating feature vectors due to dynamic analysis, they reach an accuracy of 88\%, really far from our results up to 98\%.

%% file: tex/conclusion.tex
\section{Conclusion}\label{s:conclusion}
Among the huge amount of malware daily produced, those developed by Advanced Persistent Threats (APTs) are highly relevant, as they are part of massive and dangerous campaigns that can exfiltrate information and undermine or impede critical operations of a target.

This work improves the previous automatic malware triage process, reducing the computational resources and the time required for the training phase and providing a greater flexibility. 
These improvement are achieved through the changing of the classification approach, moving from a single Multi-Class Classifier to a set of One Class classifiers.
Both algorithms were used with features obtained by static analysis on available malware known to be developed by APTs, as attested by public reports. Although static features alone are not sufficient to completely exclude relations with APTs, they allow to perform a quick triage and prioritize the analysis of malware that surely deserve higher attention, with minimal risk of wasting analysts time. In fact, the experimental evaluation has shown encouraging results: malware developed by known APTs have been detected with a precision of 100\% and an accuracy up to 95\%. 

As future work, first of all, we want to enlarge the Knowledge Base. In this work we have realized classifiers relative to 19 distinct classes, but the actual number of APTs is higher. Thus, we are interested to enriching our Knowledge Base by crawling other public APT reports, in order to be able to detect, among the huge number of malware delivered per day, the ones more relevant for further analysis.

Another interesting future development regards the improving of the classification through the revise of the \emph{features reduction} phase and the classification algorithm.  For the former, we want to observe the impact of variation of the number of projections or applying other techniques like \emph{Principal Component Analysis}; for the latter we want to test algorithms based on different principles respect to Isolation Forests, examining the possibility to rely on different approaches (not necessarily based on machine learning tools) for different APTs. This is a strong advantage of our new work, in fact we should use different classification algorithm for different APTs, choosing each time the one that performs better or combining the output of more single classifiers.

Another aspect that we want to study is the use of a different type of features: both our works use static features in order to provide a fast response, but there are many obfuscation techniques that can hide characteristics useful for the APT identification. To overcome these \emph{protections} made by adversary, it is always possible to use also dynamic analysis tools that can observe the behavior of the malware. Even if this tools require a higher amount of time respect to the static ones, we can use dynamic analysis in a second step after the first triage one, in order to validate results of the first phase or take decision on samples that have not been identified in the first step, but whose classification score is close to the threshold.